\documentstyle[aps,epsfig]{revtex}

\begin{document}

\wideabs{

\title{Spin-Momentum Correlations in Quasi-Elastic
Electron Scattering from Deuterium}

\author{
I.~Passchier$^1$,\
L.~D.~van~Buuren$^{1,2}$,\
D.~Szczerba$^3$,\
R.~Alarcon$^4$,\
Th.~S.~Bauer$^{1,5}$,\
D.~Boersma$^1$,\
J.~F.~J.~van~den~Brand$^{1,2}$,\
H.~J.~Bulten$^{1,2}$,\
R.~Ent$^{6,7}$,\
M.~Ferro-Luzzi$^{1,2}$,\
M.~Harvey$^7$,\
P.~Heimberg$^1$,\
D.~W.~Higinbotham$^{1,8}$,\
S.~Klous$^{1,2}$,\
H.~Kolster$^{1,2}$,\
J.~Lang$^3$,\
B.~L.~Militsyn$^1$,\
D.~Nikolenko$^9$,\
G.~J.~L.~Nooren$^1$,\ 
B.~E.~Norum$^8$,\ 
H.~R.~Poolman$^{1,2}$,\ 
I.~Rachek$^9$,\
M.~C.~Simani$^{1,2}$,\
E.~Six$^4$,\
H.~de~Vries$^1$,\
K.~Wang$^8$,\
Z.-L.~Zhou$^{10}$.\ \hfill}

\address{\hss\\
$^1$\ National Institute for Nuclear Physics and High Energy
      Physics, NL-1009 DB Amsterdam, The Netherlands\\
$^2$\ Faculty of Sciences, Vrije Universiteit Amsterdam, 
      NL-1081 HV Amsterdam, The Netherlands\\
$^3$\ Institut f\" ur Teilchenphysik, Eidgen\"ossische
      Technische Hochschule, CH-8093 Z\" urich, Switzerland\\
$^4$\ Department of Physics and Astronomy, 
      Arizona State University, Tempe, AZ 85287, USA\\
$^5$\ Physics Department, Utrecht University, 
      NL-3508 TA Utrecht, The Netherlands\\
$^6$\ Thomas Jefferson National Accelerator Facility, 
      Newport News, VA 23606, USA\\
$^7$\ Department of Physics, Hampton University, Hampton, VA 23668, USA\\
$^8$\ Department of Physics, University of Virginia,
      Charlottesville, VA 22901, USA\\
$^9$\ Budker Institute for Nuclear Physics, Novosibirsk, 630090
      Russian Federation\\
$^{10}$\ 
      Laboratory for Nuclear Science, Massachusetts Institute
      of Technology, Cambridge, MA 02139, USA\\
\hss\\
}

\draft
\date{\today}
\maketitle

\begin{abstract}
  We report on a measurement of spin-momentum correlations in
  quasi-elastic scattering of longitudinally polarized electrons
  with an energy of 720 MeV from
  vector-polarized deuterium.  
  The spin correlation parameter $A^V_{ed}$
  was measured for the $^2 \vec{\rm H}(\vec e,e^\prime p)n$ reaction
  for missing momenta up to 350 MeV/$c$
  at a four-momentum transfer squared of 0.21~(GeV/$c$)$^2$. The data
  give detailed information about the spin structure of the deuteron,
  and are in good agreement with the predictions of microscopic
  calculations based on realistic nucleon-nucleon
  potentials and including various spin-dependent reaction mechanism
  effects. The experiment demonstrates in a most direct manner 
  the effects of
  the $D$-state in the deuteron ground-state wave function and shows
  the importance of isobar configurations for this reaction. 
\end{abstract}


} 


The deuteron serves as a benchmark for testing nuclear theory.
Observables such as its binding energy, static magnetic dipole
and charge quadrupole moment, asymptotic ${\sl D/S}$ ratio, and the 
elastic electromagnetic form factors place strong constraints
on any realistic nuclear model.
Its simple structure allows reliable calculations to be performed
in both non-relativistic and relativistic frameworks 
\cite{Gross,Tjon,Jeschonnek,Mosconi,Forest,Arenh97}.
Such calculations are based upon state-of-the-art nucleon-nucleon (NN)
potentials \cite{Bonn,Nijmegen,Paris,av18}, and the resulting
ground-state wave function is dominated
by the $S$-state, especially at low relative proton-neutron
momentum ${\bf p}$ in the center of mass system.
Due to the tensor part of the NN interaction
a $D$-state component is generated (see {\sl e.g.} \cite{Forest,Ericson}). 
The models predict
that the $S$- and $D$-state components strongly depend 
on ${\bf p}$ and are sensitive to the repulsive core of the NN 
interaction at short distances \cite{Forest}.

Traditionally, the spin structure of the deuteron has been
studied through measurements
of the tensor analyzing power $T_{20}$ 
\cite{t20Novo1,t20Novo,t20Bates1,t20Bates,Ferro,Bouwhuis98,t20HallC}
in elastic electron-deuteron scattering.
However, more direct access to the nucleon momentum densities
is obtained by electrodisintegration studies in the region of
quasi-elastic scattering.
In the $^2{\rm H}(e,e^\prime p)n$ reaction, energy $\nu$ and 
three-momentum ${\bf q}$ are transferred to the nucleus and
the nuclear response can be mapped as a function of missing 
momentum ${\bf p}_m$ and missing energy.
Here, ${\bf p}_m \equiv {\bf q}-{\bf p}_f$ and ${\bf p}_f$ represents 
the momentum of the ejected proton. 
In this way the $(e,e^\prime p)$ reaction has been employed to
probe the proton inside the deuteron for momenta up to 1.0 GeV/$c$ 
\cite{Bernheim,Turk,Blomqvist}.
In the plane-wave impulse approximation (PWIA) 
the neutron is only a spectator during the scattering process,
and ${\bf p}_m$ is equal to the initial
proton momentum in the deuteron, 
while the missing energy equals the binding energy.

To enhance the sensitivity to the spin structure of the deuteron, 
spin dependent observables in quasi-elastic scattering can be used
\cite{Forest,ALT92,tensoreep}.
The polarization of a proton $P_z^p$ inside a 
deuteron with a vector polarization $P_1^d$, is given by \cite{ALT88}
\begin{equation}
  P_z^p = \displaystyle\sqrt{{{2}\over {3}}}
          P_1^d \left(P_S-{{1}\over{2}}P_D\right),
  \label{eq:pzvec}
\end{equation}
where $P_S$ and $P_D$ respectively represent the $S$- and $D$-state 
probability densities of the ground-state wave function. 
Note that the polarization of a nucleon in the $D$-state is opposite
to that of a nucleon in the $S$-state.

The cross section for the $^2 \vec{\rm H}(\vec e,e^\prime p)n$
reaction, in which longitudinally polarized
electrons are scattered from a polarized deuterium target,
can be written as~\cite{ALT92}
\begin{eqnarray}
S&=&S_0 ~ \{ ~ 1 ~ + ~ P_1^d ~ A^V_d ~ + ~ P_2^d ~ A^T_d \nonumber\\
    &+& h~ (~A_e ~ + ~ P_1^d ~ A^V_{ed} ~ + ~ P_2^d ~ A^T_{ed}~)~ \} \, ,
\end{eqnarray}
where $S_0$ represents the unpolarized cross section, 
$h$ the polarization of the electrons, 
and $P_1^d$ ($P_2^d$) the vector (tensor) polarization
of the target. The beam analyzing power is denoted
by $A_e$, with $A^{V/T}_d$ and $A^{V/T}_{ed}$ the
vector and tensor analyzing powers and spin-correlation parameters, 
respectively. These target analyzing powers and
spin-correlation parameters depend on the orientation of the target
spin, ${\sl e.g.}$ $A^{V/T}_{ed}(\theta_d , \phi_d )$. 
The angles $\theta_d$ and $\phi_d$ define the polarization 
direction of the deuteron in the frame where the $z$-axis is
along the direction of $\bf{q}$ and the
$y$-axis is defined by the cross product, ${\bf k} \times {\bf k^\prime}$,
of the incoming and outgoing electron momenta as shown in Fig. 1.  

\begin{figure}[htb]\unitlength1cm
\begin{picture}({8},{5.7})
\put(+0.2,0.5){\epsfxsize=7.5cm \epsfbox{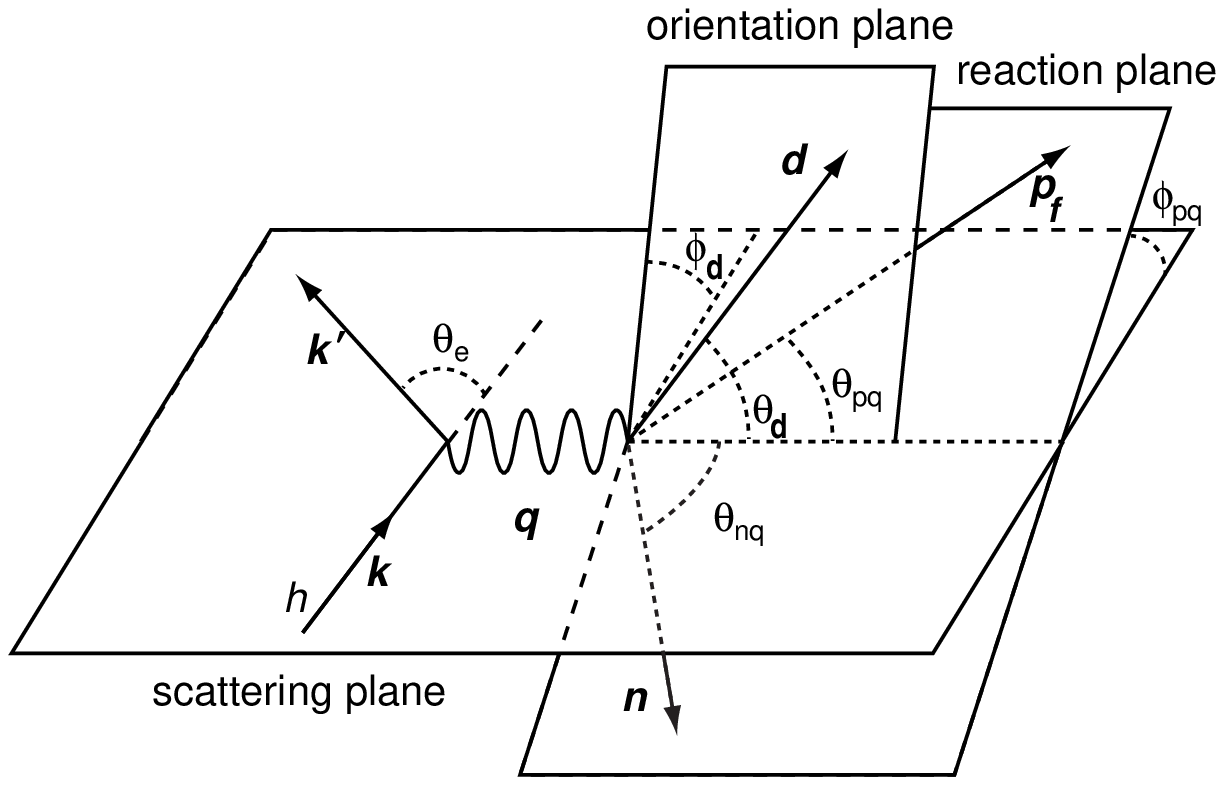}}
\end{picture}
\caption{Scattering kinematics for quasi-elastic polarized electron
  scattering from vector polarized deuterium. The target spin vector
  is represented by ${\bf d}$, while ${\bf n}$ represents the neutron.}
\label{fig:scat}
\end{figure}

In PWIA the asymmetry $A^{V}_{ed}$ in the cross section
only depends on the polarization of the proton in the deuteron given in 
Eq.~(\ref{eq:pzvec}), the kinematics of the scattering process 
and on the electromagnetic form factors of the proton 
\cite{Donnelly}. 
These form factors are well known \cite{gepBates,gepHallA} 
(see also references therein) for the kinematics 
used in the present experiment. It is therefore
possible to calculate $A^{V}_{ed}$ with high precision. 
However, the naive PWIA results must be modified to include
the contributions from the neutron (plane-wave Born approximation or PWBA) 
and to account
for spin-dependent reaction mechanism effects such as 
final-state interactions (FSI), meson-exchange currents (MEC) 
and isobar configurations (IC), while
relativistic corrections (RC) need to be applied
\cite{Arenh97}. In this letter, we report on the first
measurement of $A^{V}_{ed}$ in the 
$^2 \vec{\rm H}(\vec e,e^\prime p)n$ reaction.

The experiment was performed with a polarized gas target internal to
the Amsterdam Pulse-Stretcher (AmPS) electron storage ring\cite{Lui96}.  
Polarized electrons were produced by photo-emission from a
strained-layer semiconductor cathode (InGaAsP)\cite{bol96},
accelerated to 720~MeV, and injected  in the AmPS storage ring. 
By injecting multiple electron bunches into the storage ring,
beam currents of more than 100~mA with a life time 
in excess of 15 minutes were obtained.  
The polarization of the stored electrons was maintained by 
setting the spin tune to 0.5 with a strong solenoidal field, 
using the Siberian snake principle \cite{DK73} 
and was monitored regularly by using laser back-scattering \cite{IgorPol}. 
In order to avoid a systematic uncertainty
associated with possible beam polarization losses and to maintain a 
high average beam current, the stored electrons were dumped every 5 minutes, 
and the ring was refilled after reversal of the electron
polarization at the source.  

An atomic beam source
(ABS) produced a flux of $3\times 10^{16}$ deuterium atoms/s
in two hyperfine states\cite{Laurens}. These polarized atoms, 
analyzed by a Breit-Rabi polarimeter \cite{Laurens},
were fed into a cylindrical storage cell cooled to 75~K. 
The cell had a diameter of 15~mm and was 60~cm
long, resulting in a typical target thickness of $1 \times 10^{14}$
deuterons/cm$^2$.  An electromagnet was used to provide a guide field
of 40~mT over the storage cell. 
In order to measure $A^V_{ed}$(90$^\circ$, 0$^\circ$), the deuteron
polarization axis was oriented in the scattering plane 
and perpendicular to the $\bf{q}$ direction. 
The vector polarization of the target, $P^d_1 = \sqrt{\frac{3}{2}}
(n_+ -n_-)$, with $n_\pm$ the fraction of deuterons with spin
projection $\pm 1$, was varied every 10 seconds, while keeping
the tensor polarization fixed.

\begin{figure}[htb]\unitlength1cm
\begin{picture}({8},{7.8})
\put(+0.2,0.5){\epsfxsize=7.5cm \epsfbox{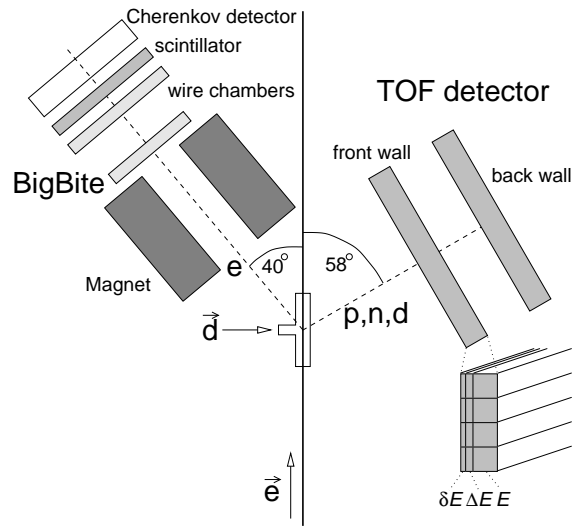}}
\end{picture}
\caption{Layout of the detector setup. The electron
  spectrometer consists of a 1~T$\cdot$m magnet, two multi-wire drift 
  chambers, a scintillator and a \v Cerenkov detector.  
  The time-of-flight system consists of two identical walls of four
  $E$-scintillators preceded by two ($\delta E$ and $\Delta E$) veto
  scintillators. The second wall was used only for neutron detection,
  as described in Ref. \protect\cite{Passchier99}.}
\label{fig:setup}
\end{figure}

Scattered electrons were detected in the large-acceptance magnetic
spectrometer BigBite\cite{Lange98}
with a momentum acceptance from 250 to
720~MeV/$c$ and a solid angle of 96~msr as shown in Fig.~\ref{fig:setup}.
BigBite was positioned at a central scattering angle of 40$^\circ$, 
resulting in a central value of 
$Q^2 \equiv {\bf q}^2 - \nu^2 = 0.21~({\rm GeV}/c)^2$.

Knocked-out protons were detected in a time-of-flight (TOF) system made of
a scintillator array, consisting of four 160~cm long, 20~cm high, and
20~cm thick vertically stacked plastic scintillator bars.  Each bar
was preceded by two ($\delta E$ and $\Delta E$) plastic scintillators
(3 and 10~mm thick, respectively) of equal length and height, used for
particle identification. Each of the scintillators was read out at
both ends to obtain position information along the bars (resolution
$\sim 4$~cm) and good coincidence timing resolution ($\sim 0.5$~ns).
The TOF detector was positioned at a central angle of 58$^\circ$ and
covered a solid angle of about 250~msr.  

Protons with kinetic energies
in excess of 40~MeV were detected with an energy resolution of about
10\%.  The $e^\prime p$ trigger was formed by a coincidence between
the electron arm trigger and a hit in any one of the TOF bars.
Protons were selected by a valid hit in two photomultipliers (PMTs) 
of at least one $E$-bar and a valid hit in both PMTs of one of 
the preceding $\Delta E$ bars. 
This requirement allowed us to use $\Delta E$-$E$ particle
identification to discriminate between
protons and either deuterons or pions. To select the two-body breakup, 
the electron energy was required to be larger than 450~MeV with a
reconstructed missing energy between $-$50~MeV and 50~MeV. 
Note that missing energy is defined as $E_m \equiv \nu-T_p-T_n$,
where $T_p$ and $T_n$ represent the kinetic energies of the ejected 
proton and recoiling neutron, respectively.
These requirements resulted in clean two-body breakup events, with
only a small dilution due to cell-wall events. 

The spin correlation parameter $A^V_{ed}$(90$^\circ$, 0$^\circ$)
was extracted from the measured asymmetry via
\begin{equation}
A_{exp}=\displaystyle\frac{N_{++} + N_{--} - N_{+-} - N_{-+}}
                          {N_{++} + N_{--} + N_{+-} + N_{-+}}
= hP_1^d~A^V_{ed} \, ,
\end{equation}
where $N_{\pm\pm}$ represent the number of events that pass the selection
criteria, with $h$ and $P_1^d$ either positive or negative, normalized to
the integrated luminosity in that configuration. 
The contribution 
of electrons scattering from the cell wall has been taken into account
by subtracting the normalized rate of cell-wall events from 
the observed number of events. 
We have studied the cell-wall contribution by measuring with an empty
storage cell.  The background contribution amounted to 5\% for low
missing momenta, increasing to about 40\% for $p_m=400$~MeV/$c$.
A possible dependence on the target density was investigated by
injecting various fluxes of unpolarized hydrogen into the cell and
measuring quasi-elastic nucleon knock-out events.  The target density
dependence was found to be negligible at ABS operating conditions.
Finite-acceptance effects were taken into account 
from the results of a Monte Carlo code that interpolated the 
model predictions in a dense grid over the full kinematical range 
and detector acceptance.

\begin{figure}[htb]\unitlength1cm
\begin{picture}({8},{8.0})
\put(+0.2,0.5){\epsfxsize=7.5cm \epsfbox{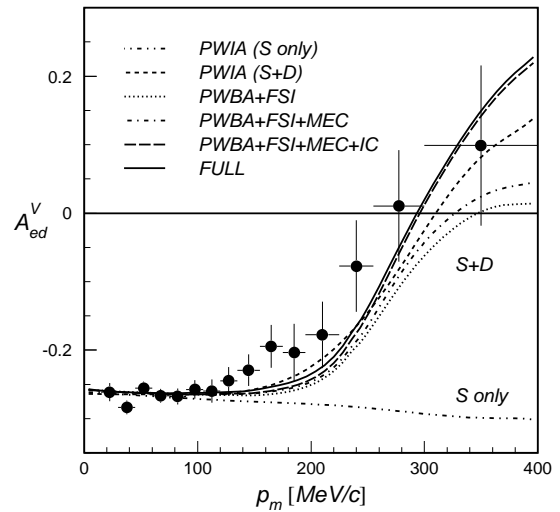}}
\end{picture}
\caption{Spin correlation parameter $A^V_{ed}$(90$^\circ$, 0$^\circ$) 
  as function of missing momentum for 
  the $^2 \vec{\rm H}(\vec e,e^\prime p)n$
  reaction at $Q^2 = 0.21~({\rm GeV}/c)^2$. 
  The short-dashed and dot-\-dot-\-dashed curves are
  PWIA predictions with and without inclusion of the $D$-wave,
  respectively. The other curves are predictions of the model of
  Arenh\"ovel \emph{et al.}, for PWBA+FSI (dotted), PWBA+FSI+MEC
  (dashed-dotted), PWBA+FSI+MEC+IC (long-dashed) and FULL 
  calculations which include RC (solid), 
  as indicated in Ref.~\protect\cite{Arenh97,ALT92}. The
  predictions are folded over the detector acceptance by 
  using a Monte Carlo method.}
\label{fig:avecbeam}
\end{figure}

Fig.~\ref{fig:avecbeam} shows the
experimental results in comparison to various predictions.
The short-dashed and dot-dot-dashed curves are PWIA predictions 
for the Argonne ${\sl v}_{18}$ NN potential 
\cite{av18} with and without inclusion 
of the $D$-state, respectively. The figure shows that inclusion
of the $D$-state is essential to obtain a fair description of the
data for the higher missing momenta.
The other curves are predictions of the model of 
Arenh\"ovel \emph{et al.} \cite{Arenh97,ALT92}
for the Bonn NN potential \cite{Bonn} and with different
descriptions for the spin-dependent reaction mechanism.
We have investigated the
dependence of the predictions on the NN potential for the
Bonn\cite{Bonn},  Nijmegen\cite{Nijmegen}, Paris\cite{Paris}
and Argonne\cite{av18} potentials. 
The effect of these potentials 
on $A^V_{ed}$ is negligible for $p_m < 200$~MeV/$c$, 
and increases to 0.04 for $p_m = 400$~MeV/$c$, 
much smaller than the accuracy of the
data or the uncertainty in the calculation of the
reaction mechanism effects. 

At $p_m < 100$ MeV/$c$, the theoretical results for $A^V_{ed}$ 
neither depend on the choice of the NN potentials 
nor on the models for the reaction mechanism.
This shows that in this specific kinematic region the deuteron
can be used as an effective neutron target.
Thus, these data were normalized to the calculations and yielded 
an absolute accuracy of 3\% in the determination of  $h P_1^d$
for our measurement of the charge form factor of the neutron 
\cite{Passchier99}.
For increasing missing momenta,
both the data and predictions for the asymmetry reverse
sign, as expected from Eq.~(\ref{eq:pzvec}) for an 
increasing contribution from the 
$D$-state component in the ground-state wave 
function of the deuteron. 
It can also be observed that inclusion of reaction
mechanism effects, mainly isobar configurations, 
are required for a better description of the data. 
This is in agreement with studies of
unpolarized quasi-elastic electron-deuteron scattering 
\cite{Blomqvist,Kasdorp98,Pellegrino,oops}.

In the region of $p_m$ around 200 MeV/$c$ where the
$S$- and $D$-states strongly interfere,
the data suggest that all models underestimate $A^V_{ed}$. 
This may be attributed to an underestimate of the $D$-state contribution
or to a lack in our understanding of the effects of $\Delta$-excitation.
This observation may be related to the deficiency in the prediction of 
the deuteron quadrupole moment by modern NN potentials
\cite{Bonn,Nijmegen,Paris,av18,Ericson}.
A similar deficit was observed in our measurements of 
$T_{20}$~\cite{Bouwhuis98} 
(see also Fig. 11 in Ref. \cite{VanOrden}). 

In summary, we have presented, for the first time, data on the spin
correlation parameter $A^V_{ed}(90^\circ,0^\circ)$ in quasi-elastic
electron-proton knock-out from the deuteron. 
The data are sensitive to the effects of the spin-dependent 
momentum distribution of the nucleons inside the deuteron. 
The experiment reveals in a most direct manner the 
effects of the $D$-state in the deuteron ground-state wave function 
and shows the importance of isobar configurations for the 
$^2 \vec{\rm H}(\vec e,e^\prime p)n$ reaction.

We would like to thank the NIKHEF and Vrije Universiteit technical
groups for their outstanding support and Prof. H. Arenh\"ovel for
providing the calculations.  This work was supported in part by the
Stichting voor Fundamenteel Onderzoek der Materie (FOM), which is
financially supported by the Nederlandse Organisatie voor
Wetenschappelijk Onderzoek (NWO), the National Science Foundation
under Grants No. PHY-9504847 (Arizona State Univ.) and No. HRD-9633750
(Hampton Univ.), US Department of
Energy under Grant No. DE-FG02-97ER41025 (Univ. of Virginia) and the
Swiss National Foundation.

\end{document}